%Paper: hep-th/9304012
%From: Ikeda Noriaki <nori@kurims.kyoto-u.ac.jp>
%Date: Mon, 5 Apr 93 16:36:38 JST

%%%%%%%%%%%%%%%%%%%%%%%%%%%%%%%%%%%%%%%%%%%%%%%%%%%%%%%%%%%%%%%%%%%%%%
%                                                                    %
%               General Form of Dilaton Gravity                      %
%                  and Nonlinear Gauge Theory                        %
%                                                                    %
%                 Noriaki Ikeda and Izawa K.-I.                      %
%                                                                    %
%                                                                    %
% This paper uses PHYZZX.TEX.                                        %
%                                                                    %
%%%%%%%%%%%%%%%%%%%%%%%%%%%%%%%%%%%%%%%%%%%%%%%%%%%%%%%%%%%%%%%%%%%%%%

\input phyzzx

\catcode`@=11
%
%%%%%%%%%%%%%% Macros for Fonts
%

%
\font\fourteentt=cmtt10 scaled\magstep2  %%% for seventeenpoint %%%
\def\seventeenpoint{\relax
    \textfont0=\seventeenrm	    \scriptfont0=\twelverm
    \scriptscriptfont0=\tenrm
     \def\rm{\fam0 \seventeenrm \f@ntkey=0 }\relax
    \textfont1=\seventeeni	    \scriptfont1=\twelvei
    \scriptscriptfont1=\teni
     \def\oldstyle{\fam1 \seventeeni\f@ntkey=1 }\relax
    \textfont2=\seventeensy	    \scriptfont2=\twelvesy
    \scriptscriptfont2=\tensy
    \textfont3=\seventeenex     \scriptfont3=\seventeenex
    \scriptscriptfont3=\seventeenex
    \def\it{\fam\itfam \seventeenit\f@ntkey=4 }
         \textfont\itfam=\seventeenit
    \def\sl{\fam\slfam \seventeensl\f@ntkey=5 }
         \textfont\slfam=\seventeensl
    \scriptfont\slfam=\twelvesl
    \def\bf{\fam\bffam \seventeenbf\f@ntkey=6 }
         \textfont\bffam=\seventeenbf
    \scriptfont\bffam=\twelvebf	 \scriptscriptfont\bffam=\tenbf
    \def\tt{\fam\ttfam \fourteentt \f@ntkey=7 }
         \textfont\ttfam=\fourteentt
    \h@big=11.9\p@{} \h@Big=16.1\p@{} \h@bigg=20.3\p@{} \h@Bigg=24.5\p@{}
    \setbox\strutbox=\hbox{\vrule height 12pt depth 5pt width\z@}
    \samef@nt}
%
%
%%%%%%%%%%%%%% Macros for the Title Page
%
\newtoks\heth
\newtoks\Heth
\Pubnum={KUNS \the\pubnum}
\Heth={HE(TH)\the\heth}
\date={\monthname,\ \number\year}
\pubnum={000}
\heth={00/00}
\def\titlepage{\FRONTPAGE\ifPhysRev\PH@SR@V\fi
   \ifp@bblock\p@bblock\fi}
\def\p@bblock{\begingroup \tabskip=\hsize minus \hsize
   \baselineskip=1.5\ht\strutbox \topspace-2\baselineskip
   \halign to\hsize{\strut ##\hfil\tabskip=0pt\crcr
   \the\Pubnum\cr \the\Heth\cr \the\date\cr }\endgroup}
\def\titlestyleb#1{\par\begingroup \interlinepenalty=9999
     \leftskip=0.00\hsize plus 1.23\hsize minus 0.02\hsize
     \rightskip=\leftskip \parfillskip=0pt
     \hyphenpenalty=9000 \exhyphenpenalty=9000
     \tolerance=9999 \pretolerance=9000
     \spaceskip=0.333em \xspaceskip=0.5em
     \iftwelv@\fourteenpoint\else\twelvepoint\fi
   \noindent {\bf #1}\par\endgroup }
\def\title#1{\vskip\frontpageskip \titlestyleb{#1} \vskip\headskip }
%
%
%%%%%%%%%%%%%% Macro for equations
%

%

%
\def\addeqno{\ifnum\equanumber<0 \global\advance\equanumber by -1
    \else \global\advance\equanumber by 1\fi}
\def\undereq#1{\mathop{\vtop{\ialign{##\crcr
      $\hfil\displaystyle{#1}\hfil$
      \crcr\noalign{\kern3\p@\nointerlineskip}
      \crcr\noalign{\kern3\p@}}}}\limits}
\def\overeq#1{\mathop{\vbox{\ialign{##\crcr\noalign{\kern3\p@}
      \crcr\noalign{\kern3\p@\nointerlineskip}
      $\hfil\displaystyle{#1}\hfil$\crcr}}}\limits}
%

%
%Equation number as you like
%

%
%%%%%%%%%%%%%% Macros for References
%
\def\journal#1&#2(#3){\unskip, {\sl #1}{\bf #2}(19#3)}
\def\andjournal#1&#2(#3){{\sl #1}{\bf #2}(19#3)}
\def\andvol&#1(#2){{\bf #1}(19#2)}

\def\NP{Nucl. Phys. }

\def\PL{Phys. Lett. }
\def\PTP{Prog. Theor. Phys. }

%
%%%%%%%%%%%%%% Macro for Acknowledgements
%

%

%%%%%%%%%%%%%%%%%%%%%%%%%%%%%%%%%%%%%%%%%%%%%%%%%%%%%%%%%%%%%%%%%%%%

\catcode`@=11 %%enables the use of @ in macros.

\def\p@bblock{\begingroup \tabskip=\hsize minus \hsize
   \baselineskip=1.5\ht\strutbox \topspace-2\baselineskip
   \halign to\hsize{\strut ##\hfil\tabskip=0pt\crcr
   \the\Pubnum\cr \the\Heth\cr \the\date\cr
   \the\pubmemo\cr %%Add a % at the top of this line for FINAL version.
   }\endgroup}

\catcode`@=12

%%%%%%%%%%%%%%%%%%%%%%%%%%%%%%%%%%%%%%%%%%%%%%%%%%%%%%%%%%%%%%%%%%%%

\footline={\hfill\ -- \folio\ -- \hfill}
\def\prenum#1{\rightline{#1}}
\def\date#1{\rightline{#1}}

%%%%%%%%%%%%%%%%%%%%%%%%%% END OF MACROS %%%%%%%%%%%%%%%%%%%%%%%%%%%%%

%%%%%%%%%%%%%%%%%%%%%%%%%%%%%%%%%%%%%%%%%%%%%%%%%%%%%%%%%%%%%%%%%%%%
%%%%%%%%%%%%%%%%%%%%%%%%%%%%%%  Refs. %%%%%%%%%%%%%%%%%%%%%%%%%%%%%%
%%%%%%%%%%%%%%%%%%%%%%%%%%%%%%%%%%%%%%%%%%%%%%%%%%%%%%%%%%%%%%%%%%%%

\REF\Iza{N.~Ikeda and Izawa K.-I.
         \journal \PTP &89 (93) to be published.}

\REF\Ike{N.~Ikeda and Izawa K.-I. \journal \PTP &89 (93) 223.}

\REF\Rus{J.G.~Russo and A.A.~Tseytlin \journal \NP &B382
	 (92) 259.}

\REF\Sch{K.~Schoutens, A.~Sevrin, and P.~van Nieuwenhuizen
         \journal Commun.~Math.~Phys. &124 (89) 87;
         \andjournal Int.~J.~Mod.~Phys. &A6 (91) 2891;
         \andjournal \PL &B255 (91) 549.}

\REF\Ver{H.~Verlinde, in {\sl String Theory and Quantum Gravity '91},
         ed.~J.~Harvey {\it et al.} (World Scientific, 1992).}

\footline={\hfil}
%\nopubblock

%\KUNS={1???}		%Please insert a KUNS number.
%\HETH={93/??}		%Please insert a HE/TH number.

%\titlepage

%\hskip 8.5cm
%July, 1992

\prenum{RIMS-918}
\date{April,1993}
\vskip 2.5cm

%\vskip 2.5cm

\title{General Form of Dilaton Gravity \break
       and Nonlinear Gauge Theory}

\author{Noriaki Ikeda}

\address{Research Institute for Mathematical Sciences \break
            Kyoto University, Kyoto 606, Japan}

\andauthor{Izawa K.-I.}
\address{Department of Physics, Kyoto University \break
                    Kyoto 606, Japan}

\abstract{
We construct a gauge theory
based on general nonlinear Lie algebras.
The generic form of `dilaton' gravity
is derived from nonlinear Poincar{\' e} algebra,
which exhibits a gauge-theoretical origin
of the non-geometric scalar field
in two-dimensional gravitation theory.
}

\endpage
\pageno=2
\footline{\hfill-\ \folio \ -\hfill}

%\sequentialequations

%\doublespace

%%%%%%%%%%%%%%%%%%%%%%%%%%%%%%%%%%%%%%%%%%%%%%%%%%%%%%%%%%%%%%%%%%%%
%%%%%%%%%%%%%%%%%%%%%%%%%%%%%%  Defs. %%%%%%%%%%%%%%%%%%%%%%%%%%%%%%
%%%%%%%%%%%%%%%%%%%%%%%%%%%%%%%%%%%%%%%%%%%%%%%%%%%%%%%%%%%%%%%%%%%%

\font\sc=cmr5 scaled\magstep1
\def\gct{\delta_{\hbox {\sc G}}}
\def\iso{\delta}
\def\brs{\delta_{\hbox {\sc B}}}
\def\brso{\delta_{\hbox {\sc b}}}

%%%%%%%%%%%%%%%%%%%%%%%%%%%%%%%%%%%%%%%%%%%%%%%%%%%%%%%%%%%%%%%%%%%%
%%%%%%%%%%%%%%%%%%%%%%%%%%%%%%%%%%%%%%%%%%%%%%%%%%%%%%%%%%%%%%%%%%%%
%%%%%%%%%%%%%%%%%%%%%%%%%%%%%% Chp. 1 %%%%%%%%%%%%%%%%%%%%%%%%%%%%%%
%%%%%%%%%%%%%%%%%%%%%%%%%%%%%%%%%%%%%%%%%%%%%%%%%%%%%%%%%%%%%%%%%%%%
%%%%%%%%%%%%%%%%%%%%%%%%%%%%%%%%%%%%%%%%%%%%%%%%%%%%%%%%%%%%%%%%%%%%
%
\chapter{Introduction}

The investigation of diffeomorphism-invariant field theories
is expected to provide indispensable information for
the quest of quantum gravity.
In fact, at the classical level, the Einstein theory
--- field theory of metric tensor with general covariance ---
offers a promising way of describing gravitational phenomena.

The general covariance in gravitation theory
is external gauge symmetry,
which may be called gauge symmetry of the Utiyama type.
Its connection to gauge symmetry of the Yang-Mills type
has been investigated on various occasions.
One of the reasons
for investigating the connection between the two types of symmetries
is that symmetry of the Yang-Mills type might be easier
to treat in quantization problems than that of the Utiyama type.

In a previous paper\rlap,
\refmark{\Iza}
we constructed
a two-dimensional gauge theory based on
quadratically nonlinear extension of
Lie algebras as a generalization
of the usual nonabelian gauge theory with internal gauge symmetry.
When the nonlinear algebra is Lorentz-covariant quadratic extension of
the Poincar\'e algebra, the theory turns out to be
the Yang-Mills-like formulation of
$R^2$ gravity with dynamical torsion\rlap.
\refmark{\Ike}
Hence it yields a new example of the connection between the
two types of gauge symmetries.

In this paper, we extend our formalism to
the case of nonlinear Lie algebras
beyond quadratic restriction made in Ref.[\Iza].
That is, we construct a gauge theory
based on general nonlinear extension of Lie algebras.
Then we also consider
the Poincar\'e algebra as a base algebra for the theory
to provide
Yang-Mills-like formulation of `dilaton' gravity.

The remaining part of the paper is organized as follows:
The next chapter provides the generic form
of `dilaton' gravity in a convenient fashion for our purposes.
In chapter 3, our action for `nonlinear' gauge theory is constructed
in quite a general manner. We also consider nonlinear Lie algebras
as a background structure for the `nonlinear' gauge theory.
The relevance of this `nonlinear' gauge theory to the `dilaton' gravity
is clarified in chapter 4.
Chapter 5 concludes the paper.
In the Appendix, we compare the above results with the formulas obtained
in our previous paper
\refmark{\Iza}
that deals with quadratically `nonlinear' gauge theory.

%%%%%%%%%%%%%%%%%%%%%%%%%%%%%%%%%%%%%%%%%%%%%%%%%%%%%%%%%%%%%%%%%%%
%%%%%%%%%%%%%%%%%%%%%%%%%%%%%%%%%%%%%%%%%%%%%%%%%%%%%%%%%%%%%%%%%%%
%%%%%%%%%%%%%%%%%%%%%%%% Chp. 2 %%%%%%%%%%%%%%%%%%%%%%%%%%%%%%%%%%%
%%%%%%%%%%%%%%%%%%%%%%%%%%%%%%%%%%%%%%%%%%%%%%%%%%%%%%%%%%%%%%%%%%%
%%%%%%%%%%%%%%%%%%%%%%%%%%%%%%%%%%%%%%%%%%%%%%%%%%%%%%%%%%%%%%%%%%%
%
\chapter{Generic Form of `Dilaton' Gravity}

The general expression for the action of two-dimensional metric
${\bar g}_{\mu \nu}$ coupled to a scalar field $\bar{\varphi}$
(without higher-derivative terms)
is given by
$$
  S = \int \! d^2 x \, \sqrt{- \bar g} \,( {1 \over 2} \bar g^{\mu\nu}
       \partial_\mu \bar{\varphi}  \partial_\nu \bar{\varphi}
	+ {1 \over 2}\, {\cal U}(\bar{\varphi} ) \bar R
         - {\cal V}(\bar{\varphi})).
 \eqn\GDGR
$$
We may
replace ${\cal U}({\bar \varphi})$ by a linear function
\refmark{\Rus}
$$
  S = \int \! d^2 x \, \sqrt{-g}
    \,({\alpha \over 2} g^{\mu\nu} \partial_\mu \varphi
     \partial_\nu \varphi + {1 \over 2}\varphi
      R - {\cal W}(\varphi) )
 \eqn\MIDDLE
$$
through a field redefinition
\foot{This field transformation is well-defined only locally
in the field space of ${\bar \varphi}$
where $(\partial {\cal U}/\partial {\bar \varphi}) \neq 0$.
In particular, it is inapplicable to the case of
${\cal U}({\bar \varphi}) = {\rm constant}$,
where the field ${\bar \varphi}$
is regarded as a scalar matter rather than the `dilaton' field.}
$$
 \eqalign{
  & \varphi = {\cal U}(\bar{\varphi}),
   \quad g_{\mu\nu} = e^{\rho} \bar{g}_{\mu\nu},
    \quad {\cal W}(\varphi) = e^{\rho} {\cal V}(\bar{\varphi}); \cr
  & \rho( \bar{\varphi}) \equiv -{\alpha}\, {\cal U}(\bar{\varphi})
    + \int \!
  %\nolimits^{\hat{\varphi}}
   d \bar{\varphi} \,
    \biggl({\partial {\cal U}
     \over \partial \bar{\varphi}} \biggr)^{-1}, \cr
 }
 \eqn\SECACT
$$
where $\alpha$ is an arbitrary constant.
In particular, we can set $\alpha = 0$ to obtain
$$
  S =  \int \! d^2x \,{\cal L}_D; \quad
  {\cal L}_D = \sqrt{- g} \,({1 \over 2}\varphi R - {\cal W}(\varphi)).
 \eqn\ACTION
$$
The action \ACTION\ is of course invariant under
the following general coordinate transformation:
$$
 \eqalign{
  \gct g_{\mu\nu} &= -v^\lambda \partial_\lambda g_{\mu\nu}
                     -(\partial_\mu v^\lambda) g_{\lambda\nu}
                     -(\partial_\nu v^\lambda) g_{\mu\lambda}, \cr
  \gct \varphi &= -v^\lambda \partial_\lambda \varphi. \cr
 }
 \eqn\DIFFE
$$

In the following chapters, we investigate
a gauge-theoretical origin of the `dilaton' field $\varphi$
in the action \ACTION, which might be regarded as a ``pure" gravity
in two dimensions.

%%%%%%%%%%%%%%%%%%%%%%%%%%%%%%%%%%%%%%%%%%%%%%%%%%%%%%%%%%%%%%%%%%%%
%%%%%%%%%%%%%%%%%%%%%%%%%%%%%%%%%%%%%%%%%%%%%%%%%%%%%%%%%%%%%%%%%%%%
%%%%%%%%%%%%%%%%%%%%%%%%%%%%%% Chp. 3 %%%%%%%%%%%%%%%%%%%%%%%%%%%%%%
%%%%%%%%%%%%%%%%%%%%%%%%%%%%%%%%%%%%%%%%%%%%%%%%%%%%%%%%%%%%%%%%%%%%
%%%%%%%%%%%%%%%%%%%%%%%%%%%%%%%%%%%%%%%%%%%%%%%%%%%%%%%%%%%%%%%%%%%%
%
\chapter{`Nonlinear' Gauge Theory}

In this chapter, we explain a generic approach to the
construction of a `nonlinear' gauge theory
--- gauge theory based on nonlinear extension of Lie algebras.

%%%%%%%%%%%%%%%%%%%%%%%%%%%%%%%%%%%%%%%%%%%%%%%%%%%%%%%%%%%%%%%%%%%%
%%%%%%%%%%%%%%%%%%%%%%%%%%%%%% Sec. 1 %%%%%%%%%%%%%%%%%%%%%%%%%%%%%%
%%%%%%%%%%%%%%%%%%%%%%%%%%%%%%%%%%%%%%%%%%%%%%%%%%%%%%%%%%%%%%%%%%%%
%
\section{`Nonlinear' Gauge Transformation}

As is the case for
quadratically `nonlinear' gauge theory
exposed in Ref.[\Iza, \Sch],
we need a `coadjoint' scalar field $\Phi_A$ in addition to
a vector field $h_\mu^A$ for the purpose of
constructing a `nonlinear' gauge theory.
Here $A$ denotes an internal index.
We would like to consider
the following generic form of gauge transformation:
$$
 \eqalign{
  \delta h_\mu^A
   &= \partial_\mu c^A + U_{BC}^A(\Phi) h_\mu^B c^C, \cr
  \delta \Phi_A &= W_{BA}(\Phi) c^B, \cr
 }
 \eqn\GGAUGE
$$
where
$U_{BC}^A(\Phi)$ and $W_{AB}(\Phi)$ are smooth functions of the field
$\Phi_A$ to be determined below, and
$c^A$ is a gauge-transformation parameter.
In the case of usual nonabelian gauge theory,
$U^A_{BC} = f^A_{BC}$ and $W_{BA} = f^C_{BA} \Phi_C$,
where $f^C_{AB}$ denote structure constants in some Lie algebra.
Thus we assume $U^A_{BC} = - U^A_{CB}$
and $W_{BA} = - W_{AB}$.

Let us investigate the commutator algebra of these gauge
transformations
with gauge-transformation parameters $c_1$ and $c_2$.
In order to get a closed algebra
on the scalar field $\Phi_A$
$$
  [\delta (c_1), \delta (c_2)]\Phi _A = \delta (c'_3)\Phi _A,
 \eqn\FCOMM
$$
we are naturally led to impose the following condition on the functions
$W_{AB}$:
$$
  { \partial W_{[AB} \over \partial \Phi_D }
   W_{C]D} = 0,
 \eqn\JACPHI
$$
where $[ABC]$ denotes antisymmetrization
in the indices $A$, $B$, and $C$.
This assumption results in the composite parameter
$$
  c'^A_3 = {\partial W_{BC} \over \partial \Phi_A} c_1^B c_2^C.
 \eqn\COMPPP
$$
In the next section, we show that the expression \JACPHI\
can be derived as the Jacobi identity for a nonlinear Lie algebra
with structure functions $W_{AB}$.

On the other hand,
as the commutator of two transformations
on the vector field $h_\mu^A$, we obtain
$$
  [\delta (c_1), \delta (c_2)]h_\mu^A = \delta (c_3)h_\mu^A + \cdots,
 \eqn\HCOMM
$$
where the composite parameter $c_3$ is given by
$$
  c_3^A = U^A_{BC} c_1^Bc_2^C,
 \eqn\COMPP
$$
and the ellipsis on the right-hand side of \HCOMM\
indicates a term which is irrelevant here (see (3.9) below).

We require $c_3 = c'_3$, that is, we henceforth put
$$
  U^A_{BC}( \Phi) =  {\partial W_{BC} \over \partial \Phi_A}.
 \eqn\UANDW
$$
in order for the commutator algebra to represent
a consistent composition law of Lagrangian symmetry.
The resulting form of the gauge transformation
is given by
$$
 \eqalign{
  & \delta h_\mu^A
  = \partial_\mu c^A + {\partial W_{BC}(\Phi) \over \partial \Phi_A }
  h_\mu ^B c^C, \cr
  & \delta \Phi_A = W_{BA} (\Phi) c^B \cr
 }
 \eqn\APHITR
$$
in terms of the functions $W_{AB}$ which satisfy the condition \JACPHI.
This gauge transformation reduces to the usual nonabelian one
when $W_{BA}(\Phi) = f_{BA}^C \Phi_C$.

Then the gauge algebra is given by
$$
 \eqalign{
  &[\delta (c_1), \delta (c_2)]h_\mu ^A \,= \delta (c_3)h_\mu ^A
             - c_1^Cc_2^D { \partial^2 W_{CD} \over
		\partial \Phi_A \partial \Phi_B } D_\mu \Phi _B, \cr
  &[\delta (c_1), \delta (c_2)]\Phi _A = \delta (c_3)\Phi _A, \cr
 }
 \eqn\FINGA
$$
where we have defined
$$
  D_\mu \Phi _A = \partial_\mu \Phi _A + W_{AB}(\Phi) h_\mu ^B.
 \eqn\DEFCOV
$$
Let us see the gauge transform of \DEFCOV:
$$
  \delta (D_\mu \Phi _A) = (D_\mu \Phi _C)
		{ \partial W_{BA} \over \partial \Phi_C} c^B.
 \eqn\DFTRA
$$
This reveals that the object $D_\mu $ may be recognized as a covariant
differentiation, $D_\mu \Phi _A$ transforming as a coadjoint vector.

The commutator of two covariant differentiations
provides
$$
  [D_\mu , D_\nu ]\Phi _A = W_{AB}(\Phi)R_{\mu\nu}^B,
 \eqn\COMMD
$$
where the curvature $R_{\mu \nu}^A$ is defined by
$$
  R_{\mu \nu }^A = \partial_\mu h_\nu ^A - \partial_\nu h_\mu ^A
            + {\partial W_{BC} \over \partial \Phi_A} h_\mu ^B h_\nu ^C,
 \eqn\CURVAT
$$
and the covariant differentiation on $D_\mu \Phi _A$ has been
so determined
as the resultant expression $D_\nu (D_\mu \Phi _A)$ transforms
like a coadjoint field.
Unfortunately, the curvature does not transform homogeneously:
$$
  \delta R_{\mu \nu }^A = { \partial W_{BC} \over
		\partial \Phi_A} R_{\mu \nu }^B c^C
                + \left\{ (D_\mu \Phi _D) { \partial^2 W_{BC} \over
		  \partial \Phi_D \partial \Phi_A} h_\nu ^B c^C
                 - (\mu  \leftrightarrow \nu ) \right\},
 \eqn\TRACUR
$$
which seems troublesome for the construction of invariant actions.
%We note that the Bianchi identity
%$D_{[\mu }^{}R_{\nu \rho ]}^A = 0$ holds.

%%%%%%%%%%%%%%%%%%%%%%%%%%%%%%%%%%%%%%%%%%%%%%%%%%%%%%%%%%%%%%%%%%%%
%%%%%%%%%%%%%%%%%%%%%%%%%%%%%% Sec. 2 %%%%%%%%%%%%%%%%%%%%%%%%%%%%%%
%%%%%%%%%%%%%%%%%%%%%%%%%%%%%%%%%%%%%%%%%%%%%%%%%%%%%%%%%%%%%%%%%%%%
%
\section{Jacobi Identity for Nonlinear Lie Algebra}

This section is devoted to a brief explanation
of an algebraic background for the `nonlinear' gauge transformation
defined in the previous section. The relevant object is
the Jacobi identity for nonlinear Lie algebras\rlap.
\refmark{\Iza, \Sch}

Let us consider a vector space with a basis $\{ T_A \}$,
the polynomial ring
\foot{Generally speaking, it is not necessary to
restrict consideration
to the case of polynomial ring, which is commutative by definition.
However, it is enough to treat the commutative case
for the present purposes
because we make use of the structure functions $W_{AB}$ only in the form
$W_{AB}(\Phi)$ where $\Phi$ represents a set of scalar fields
$\Phi_C$ which do commute: $\Phi_C \Phi_D = \Phi_D \Phi_C$.}
of whose elements yields
a distributive algebra
with a bracket product
$$
  [T_A, T_B] = W_{AB}(T),
 \eqn\NLLALG
$$
where $W_{AB}(T)$ denotes a polynomial in $T_C$
which satisfies the antisymmetry property
$W_{AB} = -W_{BA}$:
$$
  W_{AB}(T) = W^{(0)}_{AB} + W^{(1)C}_{AB}T_C + W^{(2)CD}_{AB}T_CT_D
              + \cdots,
 \eqn\POLYN
$$
where $W^{(0)}_{AB}$, $W^{(1)C}_{AB}$, and so on
stand for structure constants
in this algebra.
We note that the zeroth-order term $W^{(0)}_{AB}$ may
be recognized as a central element
\refmark{\Iza}
in the algebra \NLLALG.
The derivation nature of the bracket
($[T_A, T_B T_C] = [T_A, T_B]T_C + T_B[T_A, T_C]$, for instance)
is also assumed.
This definition includes
the ordinary Lie algebras as a special case when $W_{AB}(T)$
is linear in $T_A$, which makes the algebra \NLLALG\ deserve the name
of nonlinear Lie algebra.

The Jacobi identity to the bracket product \NLLALG\
implies
$$
  { \partial W_{[AB} \over \partial T_D }
   W_{C]D} = 0,
 \eqn\JACT
$$
which is a generalization of the Jacobi identity
for Lie algebras.

The expression \FINGA\ can be considered as a gauge algebra
based on the nonlinear Lie algebra \NLLALG\
in the sense that the functions $W_{AB}(\Phi)$ satisfy
the same constraint identity \JACPHI\ as
the Jacobi identity \JACT\ for $W_{AB}(T)$ in the algebra \NLLALG.

%%%%%%%%%%%%%%%%%%%%%%%%%%%%%%%%%%%%%%%%%%%%%%%%%%%%%%%%%%%%%%%%%%%%
%%%%%%%%%%%%%%%%%%%%%%%%%%%%%% Sec. 3 %%%%%%%%%%%%%%%%%%%%%%%%%%%%%%
%%%%%%%%%%%%%%%%%%%%%%%%%%%%%%%%%%%%%%%%%%%%%%%%%%%%%%%%%%%%%%%%%%%%
%
\section{An Invariant Action}

We do not have a straightforward way of constructing
invariant actions under the gauge transformation \APHITR,
as noticed from the inhomogeneous form of \TRACUR\
in section 3.1.
Hence we will instead seek an action which provides
covariant equations of motion under those transformations,
and examine whether it is invariant under them.

The determination of equations of motion we employ
goes as follows:
The gauge algebra \FINGA\ should be closed,
at least, on shell so as to originate from some Lagrangian symmetry.
Hence the choice
$$
  D_\mu \Phi_A = 0
 \eqn\FEOM
$$
seems an immediate guess for appropriate equations of motion,
which are indeed covariant \DFTRA\ under the transformation
\APHITR.

Since we have the vector $h_\mu^A$ and the scalar $\Phi_A$ fields
as our basic ones, a first approximation to the desired Lagrangian
might be given by
$$
  {\cal L}_0 = \chi ^{\mu \nu }h_\mu ^AD_\nu \Phi _A,
 \eqn\GUESS
$$
where $\chi ^{\mu \nu }$ denotes an invertible constant tensor
to be determined later.
However, $D_\nu \Phi _A$ itself contains the fields $h_\mu ^A$,
and hence the variation of ${\cal L}_0$ with respect to $h_\mu ^A$
does not lead to the covariant equations of motion \FEOM.

In order to overcome this issue,
we modify the Lagrangian into
$$
  {\cal L} = {\cal L}_0
            - {1\over 2}\chi ^{\mu \nu } W_{AB} h_\mu ^A h_\nu ^B,
 \eqn\TRUE
$$
which yields the desired equations of motion \FEOM\
provided $\chi ^{\mu \nu } = -\chi ^{\nu \mu }$.
Note that this Lagrangian can be rewritten as
$$
  {\cal L} = -{1\over 2}\chi ^{\mu \nu }[ \Phi _A R_{\mu \nu }^A
          + \left(W_{BC}- \Phi_A {\partial W_{BC} \over \partial
	  \Phi_A }\right) h_\mu ^Bh_\nu^C]
 \eqn\LAGCUR
$$
up to a total derivative.

If we require Poincar{\' e} invariance of the Lagrangian,
the antisymmetric tensor $\chi ^{\mu \nu }$ should be an invariant one
$$
  \chi ^{\mu \nu } = \epsilon ^{\mu \nu },
 \eqn\EPSILON
$$
where $\epsilon ^{\mu \nu }$ denotes
the Levi-Civita tensor in two dimensions.
At this point, we are led to consider the two-dimensional situation.

In summary, we have obtained a candidate action
$$
  S = \int d^2 x \, {\cal L}; \quad
  {\cal L} = -{1\over 2}\epsilon^{\mu \nu }[ \Phi _A R_{\mu \nu }^A
            + \biggl(W_{BC}- \Phi_A
           {\partial W_{BC} \over \partial \Phi_A }\biggr)
	  h_\mu ^Bh_\nu ^C ],
 \eqn\LAGL
$$
which comes out to be diffeomorphism invariant.
The relation \JACPHI\ enables us to show
$$
  \delta {\cal L} =
	\partial_\mu [ \epsilon^{\mu\nu}
	\biggl( W_{BC} - \Phi_A {\partial W_{BC} \over \partial \Phi_A }
	  \biggr) h_\nu^B c^C],
 \eqn\LAGTR
$$
which confirms the invariance of our action \LAGL\ under
the `nonlinear' gauge transformation \APHITR.
The equations of motion which follow from \LAGL\ are given by
the following:
$$
  \epsilon ^{\mu \nu }D_\nu \Phi _A = 0,
  \quad \epsilon ^{\mu \nu }R_{\mu \nu }^A = 0,
 \eqn\EQMO
$$
which are indeed covariant
due to the transformation law \DFTRA, \TRACUR.
%Let us remark that we can also arrive at the same action \LAGL\
%by a similar procedure to the above one,
%even if we start from the other equations of motion
%$R_{\mu \nu }^A = 0$ instead of \FEOM.

%%%%%%%%%%%%%%%%%%%%%%%%%%%%%%%%%%%%%%%%%%%%%%%%%%%%%%%%%%%%%%%%%%%%
%%%%%%%%%%%%%%%%%%%%%%%%%%%%%%%%%%%%%%%%%%%%%%%%%%%%%%%%%%%%%%%%%%%%
%%%%%%%%%%%%%%%%%%%%%%%%%%%%%% Chp. 4 %%%%%%%%%%%%%%%%%%%%%%%%%%%%%%
%%%%%%%%%%%%%%%%%%%%%%%%%%%%%%%%%%%%%%%%%%%%%%%%%%%%%%%%%%%%%%%%%%%%
%%%%%%%%%%%%%%%%%%%%%%%%%%%%%%%%%%%%%%%%%%%%%%%%%%%%%%%%%%%%%%%%%%%%
%
\chapter{Nonlinear Poincar{\' e} Algebra}

The general-covariance property of our action \LAGL\
suggests that we might utilize it to construct
two-dimensional gravitation theory.
It seems a natural choice to adopt the
Poincar{\' e} algebra in two dimensions
as a base algebra on which the construction will be performed.
Since the diffeomorphism invariance is guaranteed by the definition
of \LAGL, our requirement is that the theory should be local-Lorentz
covariant. Hence we consider nonlinear extension
\foot{The following form of nonlinear Poincar{\' e} algebra
is not the most general one that preserves the Lorentz structure.
It is chosen so as to realize
the torsion-free condition as an equation of motion
in the resultant `nonlinear' gauge theory (4.3).}
of the Poincar{\' e} algebra which preserves
the Lorentz structure of the genuine Poincar{\' e} algebra:
$$
 \eqalign{
  &[ \,J, J\, ] = 0, \quad
   [ \,J, P_a ] = \epsilon_{ab} \eta^{bc} P_c, \cr
  &[ P_a, P_b ] = - \epsilon_{ab} {\cal W} (J), \cr
 }
 \eqn\MODISO
$$
where $\eta^{cd}$ is the two-dimensional Minkowski metric.
Note that the choice ${\cal W}(J) = 0$ corresponds to the original
Poincar{\' e} algebra.

We set $\{ T_A \} = \{ P_a, J \}$, that is, $T_0 = P_0$, $T_1 = P_1$,
and $T_2 = J$.
Then the structure functions defined in \NLLALG\
for the above algebra are given by
$$
 \eqalign{
  W_{2a} &= - W_{a2} = \epsilon_{ab} \eta^{bc} P_c,
   \quad W_{22} = 0, \cr
  W_{ab} & = - \epsilon_{ab} {\cal W}(J). \cr
 }
 \eqn\HATENA
$$

We also set the vector field $h_\mu^A = (e_\mu{}^a, \omega_\mu)$ and
the scalar field $\Phi_A = (\phi_a, \varphi)$ to obtain
$$
  {\cal L}_S = - {1\over 2}\epsilon ^{\mu \nu } \varphi F_{\mu \nu }
               - {1\over 2}\epsilon ^{\mu \nu } \phi_a T_{\mu \nu }{}^a
               - e {\cal W} (\varphi)
 \eqn\TOPOL
$$
as the Lagrangian \LAGL,
where we have defined
$$
 \eqalign{
  F_{\mu\nu} &\equiv \partial_\mu \omega_\nu - \partial_\nu \omega_\mu,
   \cr
  T_{\mu\nu}{}^a &\equiv \partial_\mu e_\nu{}^a
	          +\omega_\mu\epsilon ^{ab}e_{\nu b}
                   -(\mu \leftrightarrow \nu), \cr
  e &\equiv {\rm det} (e_\mu{}^a). \cr
 }
 \eqn\DEFDEF
$$
The gauge transformation law \APHITR\
now reads
$$
 \eqalign{
  \iso \omega_\mu &= \partial_\mu t
                    + \epsilon_{bc} c^b e_\mu{}^c
               {\partial {\cal W}(\varphi) \over \partial \varphi}, \cr
  \iso e_\mu{}^a &= - t \epsilon^{ab} e_{\mu b}
                 + \partial_\mu c^a + \omega_\mu \epsilon^{ab} c_b, \cr
  \iso \varphi &= \epsilon^{ab} c_a \phi_b, \cr
  \iso \phi_a &= - t \epsilon_{ab} \phi^b
                + \epsilon_{ab} c^b {\cal W}(\varphi), \cr
 }
 \eqn\MISO
$$
where we have put $c^A = (c^a, t)$.

We can actually confirm
$$
  \iso {\cal L}_S
   = -\partial_\mu [\epsilon^{\mu\nu}
   e_\nu{}^a \epsilon_{ab} c^b
   \biggl( {\cal W}
    - \varphi {\partial {\cal W} \over \partial \varphi} \biggr)],
 \eqn\ISOD
$$
which corresponds to \LAGTR\ in the previous chapter.
The equations of motion which follow from \TOPOL\ are given by
$$
 \eqalign{
  &\partial_\mu \varphi + \phi_a \epsilon^{ab} e_{\mu b} = 0, \cr
  &\partial_\mu \phi_a + \omega_\mu \epsilon_{ab} \phi^b
   - \epsilon_{ab} e_\mu{}^b {\cal W} = 0, \cr
  &{1 \over 2} \epsilon^{\mu \nu} F_{\mu \nu}
    + e{{\partial {\cal W} \over \partial \varphi}} = 0, \quad
    {1 \over 2} \epsilon^{\mu \nu} T_{\mu \nu}{}^a = 0, \cr
 }
 \eqn\EQOM
$$
which of course corresponds to \EQMO.

Note that the gravitation theory \ACTION\ is obtained
if one puts $e_\mu{}^a e_\nu{}^b \eta_{ab} = g_{\mu\nu}$ and
integrates out the fields $\phi_a$ and $\omega_\mu$
in the Lagrangian \TOPOL\ under the condition $e \neq 0$.
The reformulation \TOPOL\ reveals that the theory \ACTION\ possesses
hidden gauge symmetry \MISO\ of the Yang-Mills type.
%which reduces to the ISO(1,~1) gauge transformation
%when ${\cal W}(\varphi) = 0$.
%\hbox{constant}.
%The gauge transformations \DIFFE\ and \MISO\ are reducible.
This gauge symmetry includes diffeomorphism, as is the case
for the Yang-Mills-like formulation
of $R^2$ gravity with dynamical torsion\rlap.
\refmark{\Ike}

%%%%%%%%%%%%%%%%%%%%%%%%%%%%%%%%%%%%%%%%%%%%%%%%%%%%%%%%%%%%%%%%%%%%
%%%%%%%%%%%%%%%%%%%%%%%%%%%%%%%%%%%%%%%%%%%%%%%%%%%%%%%%%%%%%%%%%%%%
%%%%%%%%%%%%%%%%%%%%%%%%%%%%%% Chp. 5 %%%%%%%%%%%%%%%%%%%%%%%%%%%%%%
%%%%%%%%%%%%%%%%%%%%%%%%%%%%%%%%%%%%%%%%%%%%%%%%%%%%%%%%%%%%%%%%%%%%
%%%%%%%%%%%%%%%%%%%%%%%%%%%%%%%%%%%%%%%%%%%%%%%%%%%%%%%%%%%%%%%%%%%%
%
\chapter{Conclusion}

We have made a systematic
construction of the `nonlinear' gauge theory \LAGL\
with the gauge transformation \APHITR\ based on
a general nonlinear Lie algebra in a parallel way to
the procedure adopted in Ref.[\Iza].
%It is intrinsically two-dimensional and diffeomorphism invariant.

%The algebraic structure of the above `nonlinear' gauge theory
%seems clear by construction.
%The clarification of its geometric structure
%is desirable in view of the rich structure
%present in the usual nonabelian gauge theory.
%However, the gauge algebra \FINGA\ is open,
%which implies non-existence of Lie-group-like object
%corresponding to the nonlinear Lie algebra.
%That might cause difficulty in the geometrical investigation
%of the `nonlinear' gauge theory.

We also considered Lorentz-covariant nonlinear extension \MODISO\
of the Poincar{\' e} algebra in two dimensions.
The `nonlinear' gauge theory based on it turned out to be
the generic form of `dilaton' gravity,
which clarified a gauge-theoretical origin of the non-geometric
scalar field in two-dimensional gravitation theory.
We note that this theory is a generalization
of Yang-Mills-like formulation of the dilaton gravity\rlap,
\refmark{\Ver}
which corresponds to a particular choice
${\cal W}(\varphi) = {\rm constant}$.

We have restricted ourselves to the classical consideration
in this paper.
It might be interesting to compare
the theory \ACTION\ of the Utiyama type
and the one \TOPOL\ of the Yang-Mills type in the aspects of their
quantization.

%%%%%%%%%%%%%%%%%%%%%%%%%%%%%%%%%%%%%%%%%%%%%%%%%%%%%%%%%%%%%%%%%%%%
%
%\acknowledge
%

%\endpage

%%%%%%%%%%%%%%%%%%%%%%%%%%%%%%%%%%%%%%%%%%%%%%%%%%%%%%%%%%%%%%%%%%%%
%%%%%%%%%%%%%%%%%%%%%%%%%%%%%%%%%%%%%%%%%%%%%%%%%%%%%%%%%%%%%%%%%%%%
%%%%%%%%%%%%%%%%%%%%%%%%%%%%%% Appen. %%%%%%%%%%%%%%%%%%%%%%%%%%%%%%
%%%%%%%%%%%%%%%%%%%%%%%%%%%%%%%%%%%%%%%%%%%%%%%%%%%%%%%%%%%%%%%%%%%%
%%%%%%%%%%%%%%%%%%%%%%%%%%%%%%%%%%%%%%%%%%%%%%%%%%%%%%%%%%%%%%%%%%%%
%
\appendix

In this Appendix,
we provide BRS formalism of the `nonlinear' gauge theory.
We also regain the quadratically `nonlinear' gauge theory
obtained in Ref.[\Iza]
as a special case of `nonlinear' gauge theory
considered in the present paper.

%%%%%%%%%%%%%%%%%%%%%%%%%%%%%%%%%%%%%%%%%%%%%%%%%%%%%%%%%%%%%%%%%%%%
%%%%%%%%%%%%%%%%%%%%%%%%%%%%%% Sec. 1 %%%%%%%%%%%%%%%%%%%%%%%%%%%%%%
%%%%%%%%%%%%%%%%%%%%%%%%%%%%%%%%%%%%%%%%%%%%%%%%%%%%%%%%%%%%%%%%%%%%
%
\section{BRS Formalism}

Let us consider
the BRS transformation $\brs$
corresponding to the classical symmetry \APHITR.
Notice that the generator algebra
for the gauge transformation \APHITR\ is open,
as can be seen from
the gauge algebra \FINGA.

We first regard the gauge-transformation
parameter $c^A$ as fermionic FP ghost.
With the aid of the condition $\iso^2 \Phi _A = 0$,
we can uniquely
determine the transformation law for the ghost
$$
 \eqalign{
  \iso c^A &= -{1 \over 2}
               {\partial W_{BC} \over \partial \Phi_A} c^B c^C, \cr
 }
 \eqn\GFPG
$$
which satisfies $\iso^2 c^A = 0$.
However, the naive definition \APHITR\
does not satisfy the nilpotency $\iso^2 =0$ on the field
$h_\mu ^A$ due to the algebra non-closure:
$$
 \eqalign{
  \iso^2 h_\mu^A &= -c^B c^C { \partial^2 W_{BC} \over \partial \Phi_A
		    \partial \Phi_D } D_\mu \Phi_D \cr
                 &= c^B c^C { \partial^2 W_{BC} \over \partial \Phi_A
		    \partial \Phi_D }
                      \epsilon_{\mu\nu} {\delta {\cal L}
                      \over \delta h_\nu^D}. \cr
 }
 \eqn\SQUAR
$$
This prevents us from performing simple-minded BRS gauge-fixing of
the Lagrangian \LAGL.

The desired BRS transformation $\brs$ and gauge-fixed Lagrangian
${\cal L}_T$ can be obtained
as follows:
In view of the equations \SQUAR,
we introduce a gauge-fixing fermionic function $\Omega $
to provide
$$
  \brs = \iso + \brso,
 \eqn\BRS
$$
where we define
$$
  \brso h_\mu^A = -i c^B c^C  { \partial^2 W_{BC} \over \partial \Phi_A
		    \partial \Phi_D }
                     \epsilon_{\mu\nu}
                    {\delta \Omega  \over \delta h_\nu^D},
 \eqn\BRSO
$$
and $\brso = 0$ on the other fields.
The gauge-fixed Lagrangian is given by
$$
  {\cal L}_T = {\cal L} - i(\brs - {1 \over 2}\brso)\Omega .
 \eqn\QUANL
$$
The BRS transformation \BRS\ is on-shell nilpotent,
and it yields a symmetry of the Lagrangian \QUANL,
as it should be:
$$
  \brs {\cal L}_T = \iso {\cal L}
   + \partial_\mu
    \biggl(i \Phi_A c^B c^C  { \partial^2 W_{BC} \over \partial \Phi_A
		    \partial \Phi_D }
     {\delta \Omega  \over \delta h_\mu^D} \biggr),
 \eqn\BRSD
$$
where $\iso {\cal L}$ is given by \LAGTR.

%%%%%%%%%%%%%%%%%%%%%%%%%%%%%%%%%%%%%%%%%%%%%%%%%%%%%%%%%%%%%%%%%%%%
%%%%%%%%%%%%%%%%%%%%%%%%%%%%%% Sec. 2 %%%%%%%%%%%%%%%%%%%%%%%%%%%%%%
%%%%%%%%%%%%%%%%%%%%%%%%%%%%%%%%%%%%%%%%%%%%%%%%%%%%%%%%%%%%%%%%%%%%
%
\section{Quadratically `Nonlinear' Gauge Theory}

Let us consider the case of quadratically nonlinear Lie algebra
$$
 \eqalign{
  [ T_A, T_B ] &= W_{AB}(T) \cr
               &= k_{AB} + f_{AB}^C T_C + V_{AB}^{CD} T_C T_D, \cr
 }
 \eqn\ANLLA
$$
where $V_{AB}^{CD} = V_{AB}^{DC}$.
Then we obtain the following identities:
$$
 \eqalign{
  &{\partial W_{BC} \over \partial \Phi_A}
   = f_{BC}^A + 2V_{BC}^{AD} \Phi_D \equiv {\tilde f}_{BC}^A, \quad
  { \partial^2 W_{BC} \over \partial \Phi_A \partial \Phi_D }
   = 2V_{BC}^{AD}, \cr
  &W_{BC}- \Phi_A {\partial W_{BC} \over \partial \Phi_A}
   = k_{BC} - V_{BC}^{AD}\Phi_A \Phi_D. \cr
 }
 \eqn\CONCRETE
$$

By virtue of the above identities, the formulas
\APHITR\ and \LAGL, respectively, result in the gauge transformation law
$$
 \eqalign{
  & \delta h_\mu^A
  = \partial_\mu c^A + {\tilde f}_{BC}^A
  h_\mu ^B c^C, \cr
  & \delta \Phi_A = (k_{BA} + f_{BA}^C\Phi_C + V_{BA}^{CD}\Phi_C\Phi_D)
   c^B, \cr
 }
 \eqn\QAPHITR
$$
and the Lagrangian
$$
 \eqalign{
  &{\cal L} = -{1 \over 2}\epsilon^{\mu \nu }[ \Phi_A R_{\mu \nu }^A
            + (k_{BC} - V_{BC}^{AD}\Phi_A \Phi_D)
	  h_\mu ^Bh_\nu ^C ]; \cr
  &R_{\mu \nu}^A = \partial_\mu h_\nu ^A - \partial_\nu h_\mu ^A
            + {\tilde f}^A_{BC} h_\mu ^B h_\nu ^C, \cr
 }
 \eqn\QLAGL
$$
which exactly coincide with the corresponding formulas
for the quadratically `nonlinear'
gauge theory given in Ref.[\Iza].

%%%%%%%%%%%%%%%%%%%%%%%%%%%%%%%%%%%%%%%%%%%%%%%%%%%%%%%%%%%%%%%%%%%%

%\endpage

\refout

\bye